**Dynamics of hybrid magnetic skyrmion driven by spin-orbit torque in ferrimagnets**

Y. Liu[1], T. T. Liu[1], Z. P. Hou[1], D. Y. Chen[1], Z. Fan[1], M. Zeng[1], X. B. Lu[1], X. S. Gao[1],

M. H. Qin[1,*], and J. –M. Liu[1,2]

[1]*Guangdong Provincial Key Laboratory of Quantum Engineering and Quantum Materials and Institute for Advanced Materials, South China Academy of Advanced Optoelectronics, South China Normal University, Guangzhou 510006, China*

[2]*Laboratory of Solid State Microstructures, Nanjing University, Nanjing 210093, China*

**[Abstract]** Magnetic skyrmions are magnetic textures with topological protection, which are expected to be information carriers in future spintronic devices. In this work, we propose a scheme to implement hybrid magnetic skyrmions (HMS) in ferrimagnets, and we study theoretically and numerically the dynamics of the HMS driven by spin-orbit torque. It is revealed that the skyrmion Hall effect depends on the skyrmion helicity and the net angular momentum ($\delta_s$), allowing the effective modulation of the HMS motion through tuning Dzyaloshinskii-Moriya interaction and $\delta_s$. Thus, the Hall effect can be suppressed through selecting suitable materials to better control the HMS motion. Moreover, Magnus force for finite $\delta_s$ suppresses the transverse motion and enhances the longitudinal propagation, resulting in the HMS dynamics in ferrimagnets faster than that in antiferromagnets.

[*]Email: qinmh@scnu.edu.cn

# I. Introduction

Magnetic skyrmions are localized spin textures with a topological number [1–5], which have been observed in a series of chiral magnets and heavy metal/ferromagnetic films where Dzyaloshinskii-Moriya interaction (DMI) due to the broken inversion symmetry plays an important role in stabilizing the skyrmions. Specifically, the bulk DMI in chiral magnets stabilizes Bloch skyrmions (typical spin configuration is shown in Fig. 1(a)) [6], and the interfacial DMI in films favors the formation of Néel skyrmions (Fig. 1(b)) [7]. Importantly, skyrmions are relatively steady with small size owing to the topological protection [8–12], and they can be driven by electric current with a rather low density, making them exciting candidates for high-density and low-power-consuming information storage devices.

Spintronic devices based on skyrmion require precise modulation of the skyrmions, while the skyrmion Hall effect remains a key challenge [13–16]. In detail, when skyrmion is driven by spin-polarized current, it suffers a transverse Magnus force and deviates from the current direction. As a result, the skyrmion can be driven out of the track under a high current, resulting in a loss of information. To overcome this deficiency, a number of solutions including the use of antiferromagnets [7], boundary effect [17], and other similar spin structures [18] have been proposed.

Most recently, hybrid magnetic skyrmion (HMS) in ferromagnets with a structure interpolating between Néel and Bloch skyrmions stabilized by the hybrid DMI (a mixture of interfacial and bulk DMIs) or by coupling with vortex structures has been reported, which is revealed to has an enhanced mobility and a reduced skyrmion Hall effect [19–22]. Interestingly, the HMS dynamics driven by spin-orbit torque (SOT) significantly depends on the skyrmion helicity, providing another parameter in modulating the Hall effect. For example, one can tune the ratio of the interfacial and bulk DMIs through various methods such as applying voltage [23], electric field [24] and strain [25,26], and in turn modulates the skyrmion helicity and dynamics. As a matter of fact, the skyrmion Hall effect induced by SOT in ferromagnets can be eliminated through delicate tuning the skyrmion helicity, while the strong stray field and relatively slow spin dynamics are still disadvantages in applications.

In antiferromagnets, on the other hand, the Magnus forces of two-sublattices induced by spin-orbit/spin-transfer torque on Néel or Bloch skyrmion are well canceled out, and the skyrmion moves along the current direction [27,28]. Thus, besides zero stray field and fast spin dynamics, antiferromagnets also provide an important platform for achieving skyrmion long-range transmission. However, effectively detection and modulation of antiferromagnetic skyrmions are still challenging in practice due to zero net magnetization of the system. Moreover, in antiferromagnets, a strong Hall motion of HMS is also revealed when it is driven by SOT, which is different from the case of spin-transfer torque driven HMS motion with zero Hall angle [29]. This phenomenon attributes to the fact that the effective SOT depending on the helicity of the HMS generally deviates from the current direction, resulting in the HMS Hall effect.

Alternatively, ferrimagnets are suggested to combine the benefits of ultra-high working frequencies with ease of detecting and modulating [30–32], noting that ferrimagnet has a nonzero magnetic moment and ultrafast spin dynamics comparable to antiferromagnets around the angular momentum compensation temperature ($T_A$). Importantly, the net angular momentum $\delta_s$ can be adjusted through tuning temperature and ion doping, which has a significant effect on skyrmion dynamics [33–38]. For example, $\delta_s$ can effectively modulate the Magnus force induced by SOT and control the skyrmion Hall motion in ferrimagnets [18,32,39].

Naturally, it is expected that the joint action of $\delta_s$ and skyrmion helicity gives rise to interesting skyrmion dynamics in ferrimagnets, which urgently deserves to be uncovered to provide a clear scenario for skyrmion motion manipulation. On the one hand, this study uncovers new dynamic behavior, contributing to the development of spintronics. For example, our earlier work has revealed that $\delta_s$ determines the momentum onto the skyrmion in ferrimagnets imposed by the injected magnons, and consequently affects the Hall angle [12]. In some extent, $\delta_s$ probably has an important effect on the SOT-driven HMS dynamics in ferrimagnets. On the other hand, the Hall effect of the HMS in ferrimagnets can be suppressed when the Magnus force depending on $\delta_s$ suppresses the transverse motion induced by the SOT.

Thus, the HMS motion with zero Hall effect could be available for certain $\delta_s$, which is instructive for experiment and device design to achieve the straight skyrmion motion.

In this work, we study the dynamics of HMS driven by SOT in ferrimagnets using Thiele theory and numerical simulations based on solving the two coupled Landau-Lifshitz-Gilbert (LLG) equations. The dependence of the skyrmion Hall angle on $\delta_s$ and the helicity is investigated, which exhibits the gradual transition in Hall angle on $\delta_s$ and the helicity. Thus, Hall motion of the HMS can be suppressed through choosing suitable $\delta_s$, allowing one to select suitable materials to better control the skyrmion motion. Furthermore, the mobility of the HMS in ferrimagnets could be enhanced attributing to Magnus force.

## II. Theory for SOT-driven HMS dynamics in ferrimagnets

In this section, we study theoretically the HMS dynamics in ferrimagnets driven by SOT through deriving the equations of motion for a HMS based on Thiele theory. We consider a spin model composed of two unequal sublattices which are antiferromagnetically coupled, as shown in Fig. 1(d). The unit magnetization vectors of the two sublattices are $\mathbf{m}_1$ and $\mathbf{m}_2$, respectively. In the continuum approximation [40], one introduces the Néel vector $\mathbf{n} = (\mathbf{m}_1 - \mathbf{m}_2)/2$ and the magnetization vector $\mathbf{m} = (\mathbf{m}_1 + \mathbf{m}_2)/2$ to deal with the dynamic equations of ferrimagnets. Considering the SOT term, the dynamics of vectors $\mathbf{m}$ and $\mathbf{n}$ can be described by the following equation,

$$s\dot{\mathbf{m}} + \delta_s \dot{\mathbf{n}} = -(\mathbf{m} \times \mathbf{f}_m + \mathbf{n} \times \mathbf{f}_n) + s_\alpha (\mathbf{n} \times \dot{\mathbf{n}}) + 2\beta (\mathbf{n} \times \mathbf{m}_p \times \mathbf{n}), \quad (1a)$$

$$s\dot{\mathbf{n}} = -(\mathbf{n} \times \mathbf{f}_m) + \alpha \left[ \delta_s (\mathbf{n} \times \dot{\mathbf{n}}) + s(\mathbf{n} \times \dot{\mathbf{m}}) \right], \quad (1b)$$

where $s = (s_1 + s_2)$ with the angular momentum densities of the two sublattices $s_1$ and $s_2$, $\delta_s = (s_1 - s_2)$, $s_\alpha = (s_1\alpha_1 + s_2\alpha_2)$ with the damping constants $\alpha_1$ and $\alpha_2$, and $\alpha = (s_1\alpha_1 + s_2\alpha_2)/s$ is the damping coefficient. $\beta = \hbar j \theta_{SH} / et$ is the coefficient of the SOT term, where $\hbar$ is the reduced Planck constant, $e$ is the electron charge, and $t$ is the thickness of the magnetic layer. $\mathbf{m}_p$ is the polarization vector of the spin polarized current, $\theta_{SH}$ is the spin Hall angle, and $j$ is the current density. $\mathbf{f}_n = -\delta U / \delta \mathbf{n}$ and $\mathbf{f}_m = -\delta U / \delta \mathbf{m}$ denote the effective fields of $\mathbf{n}$ and $\mathbf{m}$, respectively, with the free energy density $U$. Considering the fact that $|\mathbf{m}| \ll |\mathbf{n}|$ in slowly evolving system, the dynamic equation can be represented by $\mathbf{n}$ after neglecting the weak terms:

$$\rho \mathbf{n} \times \ddot{\mathbf{n}} - \delta_s \dot{\mathbf{n}} - \mathbf{n} \times \mathbf{f}_n + s_\alpha (\mathbf{n} \times \dot{\mathbf{n}}) + 2\beta (\mathbf{n} \times \mathbf{m}_p \times \mathbf{n}) = 0, \tag{2}$$

where $\rho = s^2/a$ is the inertia coefficient.

Owing to the topological protection, the HMS behaves as a soliton or a rigid body that keeps its shape unchanged during the propagation. For convenience, the HMS motion is introduced into the dynamic equation in the form of collective coordinates $\mathbf{R}(t)$, $\mathbf{n}(\mathbf{r}, t) = \mathbf{n}(\mathbf{r} - \mathbf{R}(t))$. Projecting the dynamic equation onto the HMS translational mode, the following Thiele equation of the HMS dynamics is obtained [41]:

$$M\ddot{\mathbf{R}} - \delta_s \mathbf{G} \times \dot{\mathbf{R}} - s_\alpha \mathcal{D} \cdot \dot{\mathbf{R}} + 2\beta I R(\Theta) \mathbf{m}_p = 0, \tag{3}$$

where $M = -\rho \int (\partial_i \mathbf{n} \cdot \partial_i \mathbf{n}) dxdy$ is the effective mass of the HMS, and $\mathbf{G} = (0, 0, 4\pi Q)$ is the gyromagnetic coupling vector with the topological charge $Q$. Here, $Q$, the viscous coefficient $\mathcal{D}$, and tensor $I$ read

$$Q = (1/4\pi) \int \left[ (\partial_x \mathbf{n} \times \partial_y \mathbf{n}) \cdot \mathbf{n} \right] dxdy, \tag{4a}$$

$$\mathcal{D} = \begin{pmatrix} \int (\partial_i \mathbf{n} \cdot \partial_i \mathbf{n}) dxdy & 0 \\ 0 & \int (\partial_j \mathbf{n} \cdot \partial_j \mathbf{n}) dxdy \end{pmatrix}, \tag{4b}$$

$$I = \int \left[ r \partial_r \theta + \sin \theta(r) \cos \theta(r) \right] dr, \tag{4c}$$

where $\theta$ is the polar angle of the Néel vector in the spherical coordinate, and $r$ is the radius in the polar coordinate, as shown in Fig. 1(c). The tensor $R(\Theta)$ reads

$$R(\Theta) = \begin{pmatrix} -\sin \Theta & \cos \Theta \\ -\cos \Theta & -\sin \Theta \end{pmatrix}, \tag{5}$$

where the skyrmion helicity $\Theta = \arctan(D_b/D_i)$ is determined by the bulk DMI magnitude ($D_b$) and the interfacial DMI magnitude ($D_i$), which could be altered between 0 and $2\pi$. It is worth noting that the tensor $R(\Theta)$ and SOT are directly coupled, allowing one to modulate the SOT acting on the HMS through tuning the HMS helicity.

First, we focus on the dynamics of the HMS at $T_A$ with $\delta_s = 0$. In this case, the Magnus force denoted by the second term in Eq. (3) is eliminated, and the acceleration $\ddot{\mathbf{R}}$ can be safely ignored considering the steady motion of the HMS. When the current is polarized along the $x$-direction $\mathbf{m}_p = \mathbf{e}_x$, one obtains the velocity components of the HMS,

$$v_x = -2\beta I \sin\Theta / s_\alpha \mathcal{D}, \tag{6a}$$

$$v_y = -2\beta I \cos\Theta / s_\alpha \mathcal{D}. \tag{6b}$$

It is clearly shown that the motion of the HMS depends on its helicity. Particularly, the dependence of $v_x/v_y$ on the DMI coefficients reads:

$$v_x / v_y = \tan\Theta = D_b / D_i, \tag{7}$$

the same as that obtained through solving the LLG equation in the earlier report [29].

Subsequently, we derive the velocity of the HMS in uncompensated ferrimagnets for a finite $\delta_s$, which is given by:

$$v_x = -\beta I \left(8\pi Q \delta_s \cos\Theta + 2s_\alpha \mathcal{D} \sin\Theta\right) / \left(16\pi^2 Q^2 \delta_s^2 + s_\alpha^2 \mathcal{D}^2\right), \tag{8a}$$

$$v_y = -\beta I \left(-8\pi Q \delta_s \sin\Theta + 2s_\alpha \mathcal{D} \cos\Theta\right) / \left(16\pi^2 Q^2 \delta_s^2 + s_\alpha^2 \mathcal{D}^2\right). \tag{8b}$$

Thus, the dependence of the velocity on both $\delta_s$ and $\Theta$ is clearly demonstrated, allowing one to modulate the dynamics through tuning these parameters. Taking $\Theta = \pi/4$ as an example, the Magnus force $\sim Q\delta_s$ for negative $\delta_s$ enhances the longitudinal motion and suppresses the transverse motion, resulting in the decrease of the Hall angle $\sim v_x/v_y$. Subsequently, the HMS speed $v$ and Hall angle $\theta_H$ are derived, respectively,

$$v = \frac{2\beta I}{\sqrt{16\pi^2 Q^2 \delta_s^2 + s_\alpha^2 \mathcal{D}^2}}, \tag{9a}$$

$$\theta_H = \arctan\left(\frac{-8\pi Q \delta_s \sin\Theta + 2s_\alpha \mathcal{D} \cos\Theta}{8\pi Q \delta_s \cos\Theta + 2s_\alpha \mathcal{D} \sin\Theta}\right). \tag{9b}$$

It is demonstrated that $v$ linearly depends on the current intensity and hardly be affected by the HMS helicity. Importantly, the helicity and $\delta_s$ effectively modulates the skyrmion Hall angle, demonstrating the important role of the DMI magnitudes and $\delta_s$ in controlling the HMS dynamics in ferrimagnets. For $\tan\Theta = s_\alpha \mathcal{D}/4\pi Q\delta_s$, zero Hall angle is achieved, and the skyrmion moves straightly along the current direction with a speed of $2\beta I/(16\pi Q^2\delta_s^2 + s_\alpha^2 \mathcal{D}^2)^{1/2}$. Interestingly, $s_\alpha$ decreases as $\delta_s$ increases, which may enhance the speed of the HMS. Thus, it is suggested theoretically that the Hall motion of the HMS in ferrimagnets can be completely suppressed in accompany of the speed enhancement by tuning $\delta_s$, which definitely favors future applications.

To check these predictions, a comparison between the theoretical analysis with the numerical simulations is indispensable. In Sec. III, we introduce numerical simulations of the Heisenberg spin model, and then give the calculated results and discussion.

### III. Numerical simulations and discussion

In this section, we first introduce the simulation method based on the standard Heisenberg model through solving the LLG equation, and then give the simulated and calculated results. A brief discussion on potential application of the HMS is discussed at last.

#### A. Model and simulation method

The micromagnetic simulations are performed based on the classical Heisenberg model, and the model Hamiltonian is given by

$$H = J_{AB}\sum_{<i,j>}\mathbf{s}_i \cdot \mathbf{s}_j + J_A \sum_{<i,j>_A}\mathbf{s}_i \cdot \mathbf{s}_j + J_B \sum_{<i,j>_B}\mathbf{s}_i \cdot \mathbf{s}_j \\ + D_i^A \sum_{<i,j>_A}\left(\mathbf{u}_{ij} \times \hat{z}\right)\cdot\left(\mathbf{s}_i \times \mathbf{s}_j\right) + D_b^B \sum_{<i,j>_B}\mathbf{u}_{ij} \cdot \left(\mathbf{s}_i \times \mathbf{s}_j\right) + K\sum_i \left(\mathbf{s}_i \cdot \hat{z}\right)^2, \quad (10)$$

where $\mathbf{s}_i$ is the normalized spin at lattice site $i$, $J_{AB}$ is the antiferromagnetic interlayer interaction between sublattice $A$ and sublattice $B$, $J_A$ and $J_B$ are the ferromagnetic intra-sublattice coupling for sublattice $A$ and sublattice $B$ between the nearest neighboring spins, respectively. $D_i^A$ and $D_b^B$ are the interfacial DMI and bulk DMI coefficients for two sublattices, respectively, $\mathbf{u}_{ij}$ is the unit vector connecting two spins in sublattices, and $K$ is the anisotropy constant.

The coupling parameters are the same as those in the theoretical analysis. Actually, the stabilization of ferrimagnetic skyrmions have been experimentally reported in GdCo films [42]. Herein, we consider a GdCo/Co heterostructure to produce HMS, where the coupling between the two magnetic layers can be modulated by tuning the thickness of the spacer [43]. The interfacial DMI in Co layer is induced by coupling with a heavy metal layer which generates strong spin-orbit couplings [44], and SOT can be easily applied by injecting in-plane current into the heavy metal layer [8,15]. Without loss of generality, we set the intra-sublattice exchange stiffness $A_{\text{Gd-Gd}} = 5$ pJ/m, $A_{\text{Co-Co}} = 5$ pJ/m, the perpendicular magnetic anisotropy constant $K = 0.16$ MJ/m$^3$, the bilinear surface exchange coefficient $\sigma = -1$ mJ/m$^2$, and the DMI coefficients $D_i$ and $D_b$ vary within a reasonable range.

Then, the dynamics of the HMS is investigated by solving the LLG equation,

$$\frac{\partial \mathbf{s}_i}{\partial t} = -\gamma_i \mathbf{s}_i \times \mathbf{H}_{\text{eff},i} + \alpha_i \mathbf{s}_i \times \frac{\partial \mathbf{s}_i}{\partial t} + \frac{\gamma_i \beta}{M_i} \left( \mathbf{s}_i \times \mathbf{m}_p \times \mathbf{s}_i \right), \tag{11}$$

where $\mathbf{H}_{\text{eff},i} = M_i^{-1} \partial H / \partial \mathbf{s}_i$ is the effective field with the magnetic moment $M_i$ at site $i$, and the gyromagnetic ratio $\gamma_i = g_i \mu_B / \hbar$ with the g-factors $g_1 = 2.2$ and $g_2 = 2$. We set the spin Hall angle $\theta_{\text{SH}} = 0.2$, and the damping constants $\alpha_1 = \alpha_2 = 0.4$.

Here, the micromagnetic simulations are performed using the Object-Oriented MicroMagnetic Framework (OOMMF) with extended DMI module. We start from a discrete lattice with the size of 100 nm × 100 nm × 9 nm and cell size of 1 nm × 1 nm × 3 nm, and set the time step to $10^{-13}$ s. The used magnetic moments $M_i$ for nine different cases are shown in Table 1, corresponding to nine different $\delta_s$.

## B. Results and discussion

In Fig. 2(a), we present the simulated and theoretical calculated $v_x/v_y$ as functions of $\tan\Theta$ at $T_A$. The simulations coincide well with the theory, confirming the validity of the theory analysis. A linear dependence of $v_x/v_y$ on $\tan\Theta$ is observed for $\mathbf{m}_p = \mathbf{e}_x$, the same as that in antiferromagnets [29]. Similarly, when the polarization vector of the injected spin current is tuned from $\mathbf{e}_x$ to $\mathbf{e}_y$, the HMS moves along the direction perpendicular to the motion for $\mathbf{m}_p =$

$e_x$, and $v_y/v_x$ linearly increases with tan$\Theta$. Therefore, one can modulate the HMS motion through tuning the helicity and spin polarization.

Subsequently, the effect of the helicity on the speed of the HMS $v$ is investigated, and the corresponding results are shown in Fig. 2(b) where present the simulated $v$ as a function of current density $j$ for various $D_i/D_b$ with a fixed $|D| = \sqrt{D_i^2 + D_b^2}$. $v$ linearly increases with the increasing $j$ due to the enhancement of SOT, consistent with the earlier report. For a fixed $j$, the HMS moves at a highest speed for $D_i = D_b$, and speed down when $D_i$ and $D_b$ deviate from each other. This phenomenon can be understood from the following aspects. It is well noted that skyrmion size mainly depends on DMI and anisotropy. For two uncoupled magnetic layers, DMIs with a same magnitude stabilize the isolated skyrmions with a same size, while the sizes of the skyrmions are different for DMIs with different magnitudes. Thus, for $D_i = D_b$, the skyrmions in two sublattices are well coupled due to the same skyrmion size. However, when $D_i$ deviates from $D_b$, the interlayer antiferromagnetic coupling competes with the DMI stronger than that for $D_i = D_b$, resulting in the increase of the internal energy and the reduction of the skyrmion mobility.

In some extent, this phenomenon is analogous to the barrel effect, i.e., the size and speed of the HMS are mainly determined on the smaller one of $D_i$ and $D_b$. As a matter of fact, this qualitative explanation is confirmed in the calculated $I/\mathcal{D}$ shown in the insert of Fig. 2(b), noting that $I/\mathcal{D}$ depends on the HMS structure and determines the dynamics. $I/\mathcal{D}$ reaches its maximum for $D_i = D_b$, and decreases with the deviation between $D_i$ and $D_b$. Moreover, for a same deviation magnitude, $D_i/D_b$ = 1:4 or 4:1 as an example, the HMS have a same speed, as shown in our simulations.

Subsequently, the effect of $\delta_s$ on HMS dynamics in uncompensated ferrimagnets is numerically investigated, and the corresponding results are shown in Fig. 3. Generally, during the propagation of the HMS, Magnus force proportional to $\delta_s$ is generated via the gyrotropic coupling, while the inertia coefficient $\rho$ is decreased with the increase of $\delta_s$. As a result, the speed of the HMS slightly increases as $\delta_s$ increases, as shown in Fig. 3(a) where presents the simulated and calculated $v$ as a function of $\delta_s$ for various $j$ with $\mathbf{m}_p = \mathbf{e}_x$. Furthermore, the simulations coincide well with the theory analysis for low current density ($j \sim 6 \times 10^{11}$ A/m$^2$), while deviates from the theory for large $j$. It is noted that a deformation of the HMS may occur during the propagation with a high speed, attributing to the slight discrepancy between simulations and theory.

Importantly, a significant effect of $\delta_s$ on the HSM Hall motion is revealed, as shown in Fig. 3(b) where gives the simulated velocity components $v_x$ and $v_y$ as functions of $\delta_s$ for various $j$. Here, without loss of generality, we consider the case of $D_b = D_i$, corresponding to the helicity of $-3\pi/4$. For $\delta_s = 0$, $v_x$ equals to $v_y$, and the HMS moves along the [1, 1] direction. $v_y$ increases with the increase of $\delta_s$, while $v_x$ less depends on $\delta_s$. Thus, the Hall angle of the HMS could be modulated through tuning $\delta_s$. On the one hand, the Magnus force along the [-1, 1]/[1, -1] direction is generated during the HMS propagation for positive/negative $\delta_s$, which is enhanced as $|\delta_s|$ increases. Thus, the longitudinal/transverse motion $v_x/v_y$ is suppressed/enhanced by Magnus force with the increase of $\delta_s$. On the other hand, the parameter $s_\alpha$ also decreases, and the contribution of the dissipative term to $v_x/v_y$ is enhanced. Moreover, the dissipative term competes with the Magnus force in determining $v_x$, as demonstrated in Eq. 8, resulting in a rather stable $v_x$ for various $\delta_s$. However, both the two terms contribute to $v_y$, and $v_y$ extensively increases with the increasing $\delta_s$. As a result, the HMS Hall motion can be suppressed and even eliminated by choosing suitable $\delta_s$.

Similar effect of $\delta_s$ on the Hall motion of HMS with other helicity has been revealed, and the results are summarized in Fig. 4 where gives the simulated Hall angle in the ($D_i/D_b$, $\delta_s$) parameter plane. It is clearly shown that both the HMS helicity and $\delta_s$ can be used in modulating the Hall angle, and zero Hall angle can be achieved for these parameter values shown with the dashed line. For example, the Bloch skyrmion with the helicity $\sim -\pi/2$ stabilized by $D_b$ exhibits a Hall motion, as shown in Fig. 5(a) where gives the trajectory of the skyrmion for $\delta_s = -3.1 \times 10^{-7}$ J·s/m$^3$. The Hall motion is suppressed for introducing addition $D_i = D_b/30$ which stabilize the HMS with the helicity of -91.9°, and the HMS moves straightly along the $x$-direction, as shown in Fig. 5(b). Figs. 5(c) and 5(d) gives the trajectories of the Bloch skyrmion and HMS for $\delta_s =$

3.1×10$^{-7}$ J·s/m$^3$, respectively, which also demonstrate the suppression of Hall motion by introducing suitable $D_i$. Interestingly, the HMS moves faster than the case of negative $\delta_s$, consistent with the theory analysis in Eq. (9a). In details, the speed is enhanced up to ~20% due to the enlargement of the Magnus force, noting that Magnus force suppresses the transverse motion and enhances the longitudinal propagation. Thus, ultrafast dynamics of HMS in ferrimagnets is available in the absence of the Hall motion, confirming again the advantages of ferrimagnets in future spintronic applications.

### C. Possible applications

So far, this study unveils the important role of the skyrmion helicity and $\delta_s$ in modulating the SOT-driven dynamics of the HMS in ferrimagnets, which is helpful in guiding future experiments and device design.

Generally, most of the parameters chosen in this study are comparable to those in GdCo/Co heterostructure [42–44]. The current density is in the order of 10$^{11}$ A/m$^2$ which is the typical magnitude in experiments [8,45,46], and the DMI magnitudes are set within a reasonable range. Importantly, the ratio $D_i/D_b$ has been proven to be a core control parameter in modulating the skyrmion Hall effect, allowing one to better control the HMS dynamics through tuning the DMI coefficients in ferrimagnets. In particular, the DMI can be easily adjusted through applying voltage, electric field, or strain. Of course, the theoretically revealed HMS dynamics in ferrimagnets deserves to be checked in future experiments.

A precise manipulation of skyrmion is important for implementing skyrmion-based logic devices such as logic arrays. For the application of skyrmionic logic circuit, we show here a skyrmion diversion operation in a Y-junction, as depicted in Fig. 6. Here, a gate voltage is applied to tune the DMI [24], and the electrodes are set at the nodes to control the interfacial DMI in the local area. When the input signal is "0" with low potential, no additional DMI is generated at the node and the skyrmion will move along its original track, as shown in fig. 6(a). One tunes the input voltage to "1" with high potential, a significant skyrmion Hall motion is induced and the skyrmion moves to another track, as depicted in fig. 6(b). Moreover, each voltage node can be considered as a logical unit, and the skyrmion-based programmable arrays can be implemented by connecting nodes in a series.

## V. Conclusion

In conclusion, we have studied theoretically and numerically the dynamics of HMS in ferrimagnets driven by SOT. The dependence of the skyrmion Hall angle on the net angle momentum $\delta_s$ and the helicity is unveiled by numerical simulations, which demonstrates the gradual transition in Hall angle on $\delta_s$ and the helicity. Particularly, Hall motion of the HMS can be eliminated through tuning $\delta_s$, allowing one to select suitable materials to better control the skyrmion motion. Moreover, the HMS for finite $\delta_s$ may exhibit faster dynamics than that of antiferromagnets, attributing to the Magnus force which suppresses the transverse motion and enhances the longitudinal propagation. Thus, this work demonstrates the advantages of the ferrimagnetic hybrid skyrmions for spintronic application, which is very meaningful for experiments and device design.

## Acknowledgment

We sincerely appreciate the insightful discussions with Zhejunyu Jin and Xue Liang. The work is supported by the Natural Science Foundation of China (Grants No. U22A20117, No. 51971096, No. 92163210, and No. 51721001), the Guangdong Basic and Applied Basic Research Foundation (Grant No. 2022A1515011727), and Funding by Science and Technology Projects in Guangzhou (Grant No. 202201000008).

**References:**

[1] N. Nagaosa and Y. Tokura, *Topological Properties and Dynamics of Magnetic Skyrmions*, Nat. Nanotechnol. **8**, 899 (2013).

[2] A. Fert, V. Cros, and J. Sampaio, *Skyrmions on the Track*, Nat. Nanotechnol. **8**, 152 (2013).

[3] X. Zhang, Y. Zhou, and M. Ezawa, *High-Topological-Number Magnetic Skyrmions and Topologically Protected Dissipative Structure*, Phys. Rev. B **93**, 024415 (2016).

[4] J. J. Liang, J. H. Yu, J. Chen, M. H. Qin, M. Zeng, X. B. Lu, X. S. Gao, and J. M. Liu, *Magnetic Field Gradient Driven Dynamics of Isolated Skyrmions and Antiskyrmions in Frustrated Magnets*, New J. Phys. **20**, 053037 (2018).

[5] Z. Jin, C. Y. Meng, T. T. Liu, D. Y. Chen, Z. Fan, M. Zeng, X. B. Lu, X. S. Gao, M. H. Qin, and J. M. Liu, *Magnon-Driven Skyrmion Dynamics in Antiferromagnets: Effect of Magnon Polarization*, Phys. Rev. B **104**, 054419 (2021).

[6] S. D. Pollard, J. A. Garlow, K. W. Kim, S. Cheng, K. Cai, Y. Zhu, and H. Yang, *Bloch Chirality Induced by an Interlayer Dzyaloshinskii-Moriya Interaction in Ferromagnetic Multilayers*, Phys. Rev. Lett. **125**, 227203 (2020).

[7] J. Barker and O. A. Tretiakov, *Static and Dynamical Properties of Antiferromagnetic Skyrmions in the Presence of Applied Current and Temperature*, Phys. Rev. Lett. **116**, 147203 (2016).

[8] L. Caretta, M. Mann, F. Büttner, K. Ueda, B. Pfau, C. M. Günther, P. Hessing, A. Churikova, C. Klose, M. Schneider, D. Engel, C. Marcus, D. Bono, K. Bagschik, S. Eisebitt, and G. S. D. Beach, *Fast Current-Driven Domain Walls and Small Skyrmions in a Compensated Ferrimagnet*, Nat. Nanotechnol. **13**, 1154 (2018).

[9] R. Khoshlahni, A. Qaiumzadeh, A. Bergman, and A. Brataas, *Ultrafast Generation and Dynamics of Isolated Skyrmions in Antiferromagnetic Insulators*, Phys. Rev. B **99**, 054423 (2019).

[10] H. Wu, F. Groß, B. Dai, D. Lujan, S. A. Razavi, P. Zhang, Y. Liu, K. Sobotkiewich, J. Förster, M. Weigand, G. Schütz, X. Li, J. Gräfe, and K. L. Wang, *Ferrimagnetic Skyrmions in Topological Insulator/Ferrimagnet Heterostructures*, Adv. Mater. **32**, 2003380 (2020).

[11] H. Li, C. A. Akosa, P. Yan, Y. Wang, and Z. Cheng, *Stabilization of Skyrmions in a Nanodisk Without an External Magnetic Field*, Phys. Rev. Appl. **13**, 034046 (2020).

[12] Z. Hou, Y. Wang, X. Lan, S. Li, X. Wan, F. Meng, Y. Hu, Z. Fan, C. Feng, M. Qin, M. Zeng, X. Zhang, X. Liu, X. Fu, G. Yu, G. Zhou, Y. Zhou, W. Zhao, X. Gao, and J. ming Liu, *Controlled Switching of the Number of Skyrmions in a Magnetic Nanodot by Electric Fields*, Adv. Mater. **34**, 2107908 (2022).

[13] W. Jiang, X. Zhang, G. Yu, W. Zhang, X. Wang, M. Benjamin Jungfleisch, J. E. Pearson, X. Cheng, O. Heinonen, K. L. Wang, Y. Zhou, A. Hoffmann, and S. G. E. Te Velthuis, *Direct Observation of the Skyrmion Hall Effect*, Nat. Phys. **13**, 162 (2017).

[14] K. Litzius, I. Lemesh, B. Krüger, P. Bassirian, L. Caretta, K. Richter, F. Büttner, K. Sato, O. A. Tretiakov, J. Förster, R. M. Reeve, M. Weigand, I. Bykova, H. Stoll, G. Schütz, G. S. D. Beach, and M. Klaüi, *Skyrmion Hall Effect Revealed by Direct Time-Resolved X-Ray Microscopy*, Nat. Phys. **13**, 170 (2017).

[15] R. Juge, S. G. Je, D. D. S. Chaves, L. D. Buda-Prejbeanu, J. Peña-Garcia, J. Nath, I. M. Miron, K. G. Rana, L. Aballe, M. Foerster, F. Genuzio, T. O. Menteş, A. Locatelli, F. Maccherozzi, S. S. Dhesi, M. Belmeguenai, Y. Roussigné, S. Auffret, S. Pizzini, G. Gaudin, J. Vogel, and O. Boulle, *Current-Driven Skyrmion Dynamics and Drive-Dependent Skyrmion Hall Effect in an Ultrathin Film*, Phys. Rev. Appl. **12**, 044007 (2019).

[16] L. Liu, W. Chen, and Y. Zheng, *Current-Driven Skyrmion Motion beyond Linear Regime: Interplay between Skyrmion Transport and Deformation*, Phys. Rev. Appl. **14**, 024077 (2020).

[17] Z. R. Yan, Y. Z. Liu, Y. Guang, J. F. Feng, R. K. Lake, G. Q. Yu, and X. F. Han, *Robust Skyrmion Shift Device through Engineering the Local Exchange-Bias Field*, Phys. Rev. Appl. **14**, 044008 (2020).

[18] S. Yang, K. Wu, Y. Zhao, X. Liang, J. Xia, and Y. Zhou, *Inhibition of Skyrmion Hall Effect by a Stripe Domain Wall*, Phys. Rev. Appl. **18**, 024030 (2022).

[19] H. Z. Wu, B. F. Miao, L. Sun, D. Wu, and H. F. Ding, *Hybrid Magnetic Skyrmion*, Phys. Rev. B **95**, 174416 (2017).

[20] H. Vakili, Y. Xie, and A. W. Ghosh, *Self-Focusing Hybrid Skyrmions in Spatially Varying Canted Ferromagnetic Systems*, Phys. Rev. B **102**, 174420 (2020).

[21] K. W. Kim, K. W. Moon, N. Kerber, J. Nothhelfer, and K. Everschor-Sitte, *Asymmetric Skyrmion Hall Effect in Systems with a Hybrid Dzyaloshinskii-Moriya Interaction*, Phys. Rev. B **97**, 224427 (2018).

[22] X. Zhang, J. Xia, Y. Zhou, X. Liu, H. Zhang, and M. Ezawa, *Skyrmion Dynamics in a Frustrated Ferromagnetic Film and Current-Induced Helicity Locking-Unlocking Transition*, Nat. Commun. **8**, 1717 (2017).

[23] K. Nawaoka, S. Miwa, Y. Shiota, N. Mizuochi, and Y. Suzuki, *Voltage Induction of Interfacial Dzyaloshinskii-Moriya Interaction in Au/Fe/MgO Artificial Multilayer*, Appl. Phys. Express **8**, 063004 (2015).

[24] T. Koyama, Y. Nakatani, J. Ieda, and D. Chiba, *Electric Field Control of Magnetic Domain Wall Motion via Modulation of the Dzyaloshinskii-Moriya Interaction*, Sci. Adv. **4**, eaav0265 (2018).

[25] Y. Zhang, J. Liu, Y. Dong, S. Wu, J. Zhang, J. Wang, J. Lu, A. Rückriegel, H. Wang, R. Duine, H. Yu, Z. Luo, K. Shen, and J. Zhang, *Strain-Driven Dzyaloshinskii-Moriya Interaction for Room-Temperature Magnetic Skyrmions*, Phys. Rev. Lett. **127**, 117204 (2021).

[26] O. G. Udalov and I. S. Beloborodov, *Strain-Dependent Dzyaloshinskii-Moriya Interaction in a Ferromagnet/Heavy-Metal Bilayer*, Phys. Rev. B **102**, 134422 (2020).

[27] V. Baltz, A. Manchon, M. Tsoi, T. Moriyama, T. Ono, and Y. Tserkovnyak, *Antiferromagnetic Spintronics*, Rev. Mod. Phys. **90**, 15005 (2018).

[28] Z. Jin, T. T. Liu, W. H. Li, X. M. Zhang, Z. P. Hou, D. Y. Chen, Z. Fan, M. Zeng, X. B. Lu, X. S. Gao, M. H. Qin, and J. M. Liu, *Dynamics of Antiferromagnetic Skyrmions in the Absence or Presence of Pinning Defects*, Phys. Rev. B **102**, 054419 (2020).

[29] R. Msiska, D. R. Rodrigues, J. Leliaert, and K. Everschor-Sitte, *Nonzero Skyrmion Hall Effect in Topologically Trivial Structures*, Phys. Rev. Appl. **17**, 064015 (2022).

[30] M. Guo, H. Zhang, and R. Cheng, *Manipulating Ferrimagnets by Fields and Currents*, Phys. Rev. B **105**, 064410 (2022).

[31] T. G. H. Blank, K. A. Grishunin, E. A. Mashkovich, M. V. Logunov, A. K. Zvezdin, and A. V. Kimel, *THz-Scale Field-Induced Spin Dynamics in Ferrimagnetic Iron Garnets*, Phys. Rev. Lett. **127**, 037203 (2021).

[32] S. J. Kim, D. K. Lee, S. H. Oh, H. C. Koo, and K. J. Lee, *Theory of Spin-Torque Ferrimagnetic Resonance*, Phys. Rev. B **104**, 024405 (2021).

[33] S. K. Kim, G. S. D. Beach, K.-J. Lee, T. Ono, T. Rasing, and H. Yang, *Ferrimagnetic Spintronics*, Nat. Mater. **21**, 24 (2022).

[34] D. H. Kim, D. H. Kim, K. J. Kim, K. W. Moon, S. Yang, K. J. Lee, and S. K. Kim, *The Dynamics of a Domain Wall in Ferrimagnets Driven by Spin-Transfer Torque*, J. Magn. Magn. Mater. **514**, 167237 (2020).

[35] S. H. Oh, S. K. Kim, D. K. Lee, G. Go, K. J. Kim, T. Ono, Y. Tserkovnyak, and K. J. Lee, *Coherent Terahertz Spin-Wave Emission Associated with Ferrimagnetic Domain Wall Dynamics*, Phys. Rev. B **96**, 100407(R) (2017).

[36] K. J. Kim, S. K. S. Kim, Y. Hirata, S. H. Oh, T. Tono, D. H. Kim, T. Okuno, W. S. Ham, S. K. S. Kim, G. Go, Y. Tserkovnyak, A. Tsukamoto, T. Moriyama, K. J. Lee, and T. Ono, *Fast Domain Wall Motion in the Vicinity of the Angular Momentum Compensation Temperature of Ferrimagnets*, Nat. Mater. **16**, 1187 (2017).

[37] S. K. Kim, K. J. Lee, and Y. Tserkovnyak, *Self-Focusing Skyrmion Racetracks in Ferrimagnets*, Phys. Rev. B **95**, 140404(R) (2017).

[38] S. K. Kim, K. Nakata, D. Loss, and Y. Tserkovnyak, *Tunable Magnonic Thermal Hall Effect in Skyrmion Crystal Phases of Ferrimagnets*, Phys. Rev. Lett. **122**, 057204 (2019).

[39] A. K. C. Tan, P. Ho, J. Lourembam, L. Huang, H. K. Tan, C. J. O. Reichhardt, C. Reichhardt, and A. Soumyanarayanan, *Visualizing the Strongly Reshaped Skyrmion Hall Effect in Multilayer Wire Devices*, Nat. Commun. **12**, 6 (2021).

[40] E. G. Tveten, T. Müller, J. Linder, and A. Brataas, *Intrinsic Magnetization of Antiferromagnetic Textures*, Phys. Rev. B **93**, 104408 (2016).

[41] A. A. Thiele, *Steady-State Motion of Magnetic Domains*, Phys. Rev. Lett. **30**, 230 (1973).

[42] R. Streubel, C.-H. Lambert, N. Kent, P. Ercius, A. T.N'Diaye, C. Ophus, S. Salahuddin, and P. Fischer, *Experimental Evidence of Chiral Ferrimagnetism in Amorphous GdCo Films*, Adv. Mater. **30**, 1800199 (2018).

[43] J. Chatterjee, D. Polley, A. Pattabi, H. Jang, S. Salahuddin, and J. Bokor, *RKKY Exchange Bias Mediated Ultrafast All-Optical Switching of a Ferromagnet*, Adv. Funct. Mater. **32**, 2107490 (2022).

[44] F. Tejo, F. Velozo, R. G. Elías, and J. Escrig, *Oscillations of Skyrmion Clusters in Co/Pt Multilayer Nanodots*, Sci. Rep. **10**, 16517 (2020).

[45] S. Woo, K. M. Song, X. Zhang, Y. Zhou, M. Ezawa, X. Liu, S. Finizio, J. Raabe, N. J. Lee, S. Il Kim, S. Y. Park, Y. Kim, J. Y. Kim, D. Lee, O. Lee, J. W. Choi, B. C. Min, H. C. Koo, and J. Chang, *Current-Driven Dynamics and Inhibition of the Skyrmion Hall Effect of Ferrimagnetic Skyrmions in GdFeCo Films*, Nat. Commun. **9**, 959 (2018).

[46] S. Ghosh, T. Komori, A. Hallal, J. Peña Garcia, T. Gushi, T. Hirose, H. Mitarai, H. Okuno, J. Vogel, M. Chshiev, J. P. Attané, L. Vila, T. Suemasu, and S. Pizzini, *Current-Driven Domain Wall Dynamics in Ferrimagnetic Nickel-Doped Mn4N Films: Very Large Domain Wall Velocities and Reversal of Motion Direction across the Magnetic Compensation Point*, Nano Lett. **21**, 2580 (2021).

Table 1. Parameters used in the numerical simulation.

| Index | 1 | 2 | 3 | 4 | 5 | 6 | 7 | 8 | 9 |
|---|---|---|---|---|---|---|---|---|---|
| $M_1$ (kA/m) | 1140 | 1130 | 1120 | 1110 | 1100 | 1090 | 1080 | 1070 | 1060 |
| $M_2$ (kA/m) | 1080 | 1060 | 1040 | 1020 | 1000 | 980 | 960 | 940 | 920 |
| $\delta_s$ (×10$^{-7}$ J·s/m$^3$) | −2.48 | −1.86 | −1.24 | −0.62 | 0 | 0.62 | 1.24 | 1.86 | 2.48 |

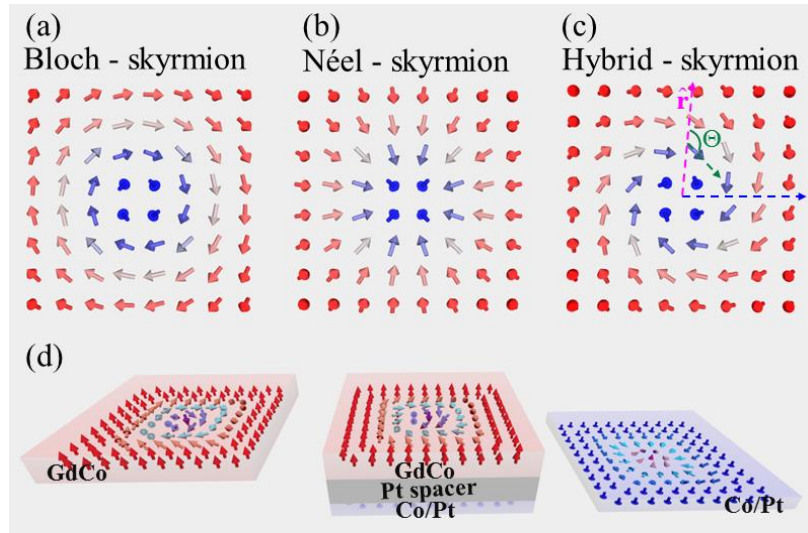

FIG. 1. Schematic spin configuration of (a) Bloch-skyrmion, (b) Néel-skyrmion and (c) hybrid skyrmion, and (d) sketch of a system with hybrid skyrmion where red and blue arrows represent the magnetization direction of CoGd and Co/Pt alloy, respectively.

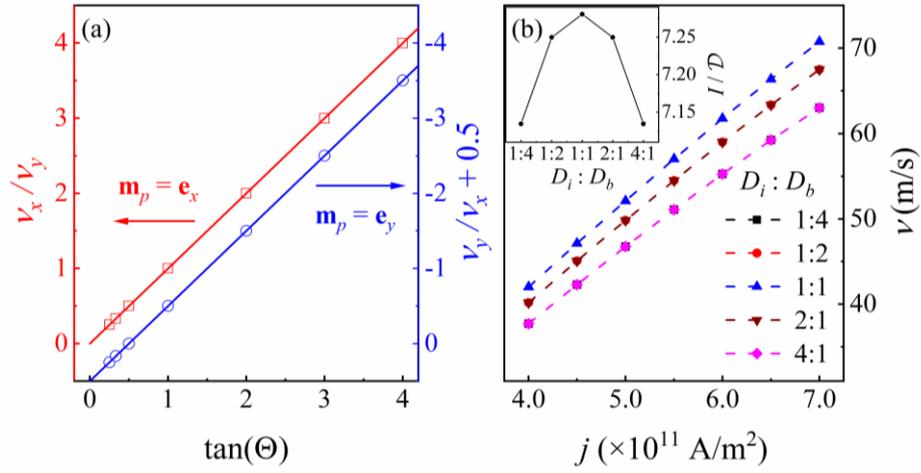

FIG. 2. (a) The calculated (lines) and simulated (symbols) $v_x/v_y$ as functions of the skyrmion helicity $\tan\Theta$ for various $\mathbf{m}_p$, and (b) the simulated HMS speed as a function of the current density $j$ for various DMI ratios. The insert shows the calculated $I/\mathcal{D}$ for $j = 5 \times 10^{11}$ A/m$^2$ for various DMI ratios.

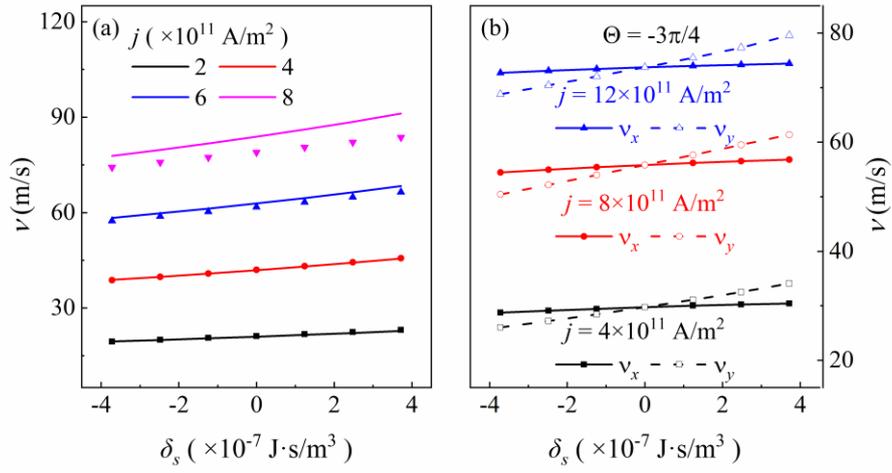

FIG. 3. (a) The calculated (lines) and simulated (symbols) $v$, and (b) the simulated $v_x$ and $v_y$ of the HMS as a function of $\delta_s$ for various $j$.

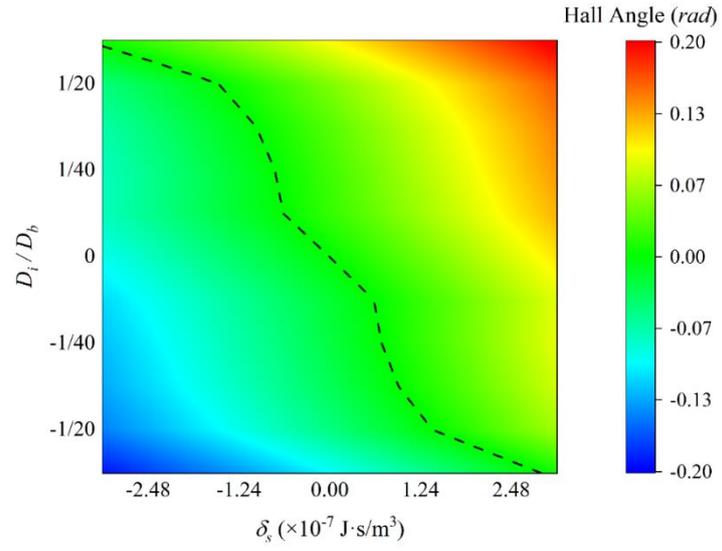

FIG. 4. The simulated Hall angle of the HMS for $\mathbf{m}_p = \mathbf{e}_x$ in the ($D_i/D_b$, $\delta_s$) parameter plane. The black dashed line corresponds to the absence of the skyrmion Hall effect.

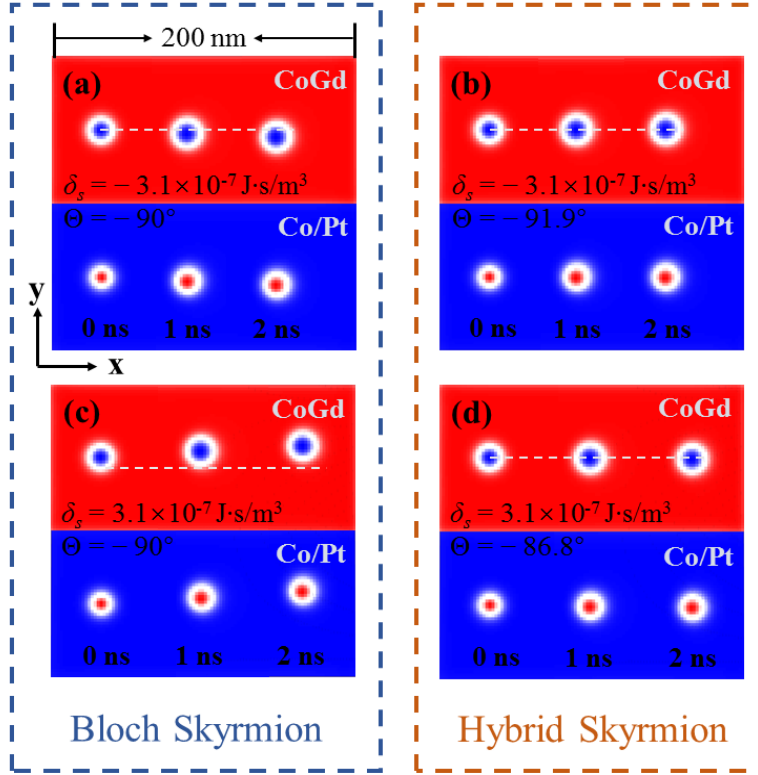

FIG. 5. Trajectories of the skyrmion for (a) $\delta_s = -3.1 \times 10^{-7}$ J·s/m$^3$ and $\Theta = -90°$, and (b) $\delta_s = -3.1 \times 10^{-7}$ J·s/m$^3$ and $\Theta = -91.9°$, and (c) $\delta_s = 3.1 \times 10^{-7}$ J·s/m$^3$ and $\Theta = -90°$, and (d) $\delta_s = 3.1 \times 10^{-7}$ J·s/m$^3$ and $\Theta = -86.8°$. The magnetic textures in (a) and (c) are Bloch-type skyrmions, while those in (b) and (d) are HMS stabilized by the additional interfacial DMI.

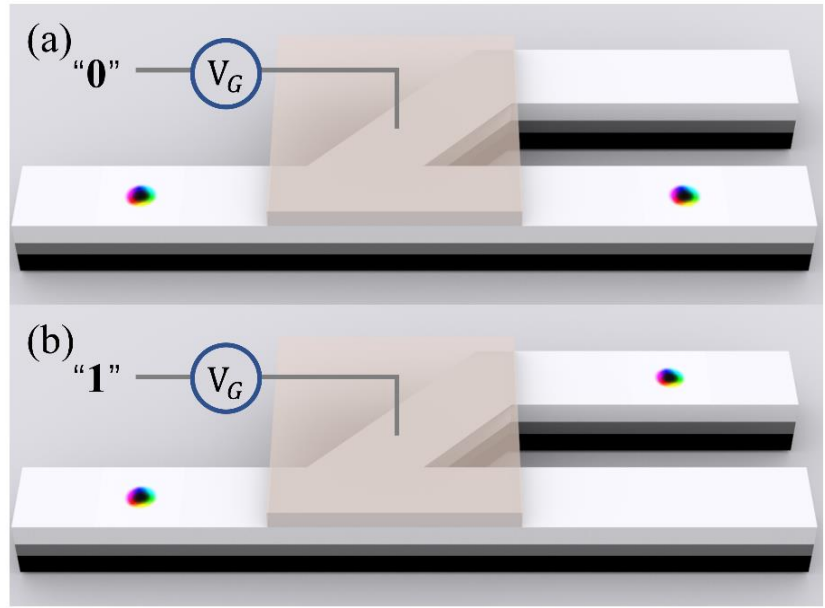

FIG. 6. The propagation of HMS driven by SOT in the Y-junctions with the input of (a) low potential "0", and (b) high potential "1". Here, the interfacial DMI is tuned by the applied voltage.